\documentclass[11pt]{article}

\usepackage{moreverb,url}

\usepackage[colorlinks,bookmarksopen,bookmarksnumbered,citecolor=red,urlcolor=red]{hyperref}

\newcommand\BibTeX{{\rmfamily B\kern-.05em \textsc{i\kern-.025em b}\kern-.08em
T\kern-.1667em\lower.7ex\hbox{E}\kern-.125emX}}

\usepackage[a4paper,right=1in,left=1in,top=1in,bottom=1in]{geometry}
\usepackage{amsmath,amssymb,amsthm}
\usepackage{graphicx}
\usepackage{subfig}

\usepackage{color, soul}
\usepackage[dvipsnames]{xcolor}
\definecolor{light-gray}{gray}{0.95}
\colorlet{laurencol}{cyan!10!light-gray}
\colorlet{laurencol2}{red!10!light-gray}
\usepackage{colortbl}

\usepackage{tikz}
\usetikzlibrary{decorations.pathreplacing,calc}
\usepackage[many]{tcolorbox}
\newtcolorbox{rightbrace}{%
    enhanced jigsaw, 
    breakable, 
    frame hidden, 
    parbox=false,
}

\title{Multiple imputation with missing data indicators}
\author{\textbf{Lauren J Beesley}$^1$, \textbf{Irina Bondarenko}$^1$, \textbf{Michael R Elliott}$^{1,2}$, \\
\textbf{Allison W Kurian}$^4$, \textbf{Steven J Katz}$^3$, \textbf{Jeremy M G Taylor}$^{*1}$\\
\\
$^1$ Department of Biostatistics, University of Michigan\\
$^2$ Survey Methodology Program, Institute for Social Research\\
$^3$ Department of Internal Medicine, University of Michigan\\
$^4$ Departments of Medicine and Epidemiology and Population Health, Stanford University\\
*Corresponding Author: jmgt@umich.edu
}
\date{\today}   

\begin{document}

\maketitle 

\allowdisplaybreaks
\raggedbottom

\begin{abstract}
Multiple imputation is a well-established general technique for analyzing data with missing values. A convenient way to implement multiple imputation is sequential regression multiple imputation (SRMI), also called chained equations multiple imputation. In this approach, we impute missing values using regression models for each variable, conditional on the other variables in the data. This approach, however, assumes that the missingness mechanism is missing at random, and it is not well-justified under not-at-random missingness without additional modification. In this paper, we describe how we can generalize the SRMI imputation procedure to handle not-at-random missingness (MNAR) in the setting where missingness may depend on other variables that are also missing. We provide algebraic justification for several generalizations of standard SRMI using Taylor series and other approximations of the target imputation distribution under MNAR. Resulting regression model approximations include indicators for missingness, interactions, or other functions of the MNAR missingness model and observed data. In a simulation study, we demonstrate that the proposed SRMI modifications result in reduced bias in the final analysis compared to standard SRMI, with an approximation strategy involving inclusion of an offset in the imputation model performing the best overall. The method is illustrated in a breast cancer study, where the goal is to estimate the prevalence of a specific genetic pathogenic variant.
\end{abstract}

Keywords: chained equations multiple imputation, not missing at random, missing data indicator, sequential regression multiple imputation

\section{Introduction}

Multiple imputation has become a popular and effective approach for analyzing datasets with missing values \cite{Rubin1987, Little2002, White2011}. This general approach relies on an assumed statistical model for the variables with missing values. If this model is appropriately specified and the mechanism generating missingness in the data depends only on fully-observed data (called missing at random [MAR]), then this method has been shown to have good theoretical and numerical properties \cite{Carpenter2013}. In analyzing data in practice, analysts must make good choices in specifying models used for imputation, and they must determine whether the MAR missingness assumption is plausible or at least approximately satisfied.

When missingness depends on unobserved data conditional on the observed data, called missing not at random (MNAR), then many standard multiple imputation strategies cannot be directly applied \cite{Little2002}. For example, suppose we have three variables in our data (denoted $X_1$, $X_2$, and $X_3$) and that $X_1$ and $X_2$ have missing values for some subjects. Let $R_j$ be the indicator of whether $X_j$ is observed ($R_j = 1$) or not ($R_j = 0$). Missingness in $X_1$ is MAR if $P(R_1 = 1 \vert X)$ depends only on $X_3$. Missingness is MNAR if missingness in $X_1$ depends directly on the value of $X_1$ or if it depends on $X_2$, which is also sometimes missing. If we were to impute missing values of $X_1$ and $X_2$ ignoring MNAR missingness, we may introduce bias in estimating parameters of interest later on. 

It is well-known that it is difficult or impossible to distinguish between MNAR and MAR missingness using the observed data alone \cite{Molenberghs2008}. Therefore, a general recommendation is to use a large number of observed variables to impute the missing data, since it may be more reasonable to assume MAR missingness when we condition on a larger amount of the observed data. Another general approach is to perform a sensitivity analysis exploring how much final analysis conclusions are impacted when we perform imputation from distributions incorporating different plausible MNAR assumptions (i.e., models for $R_1$ and $R_2$ with corresponding fixed parameter values). These imputation distributions, however, can often be complicated functions of the data models (models for $X$) and the assumed models for missingness (model for $R \vert X$). Approximations of these imputation distributions can provide an easier path toward routine implementation.

The ideal way to impute variables with missing values under MAR is to specify a joint distribution for all the $X$ variables and then use the conditional distribution derived from that joint distribution to impute missing values. It is challenging to specify such a joint distribution when many variables have missing values and the variables may be of mixed types, such as binary, categorical and continuous. A convenient and pragmatic way to overcome this problem is to perform chained equations multiple imputation, also known as sequential regression multiple imputation [denoted SRMI] \cite{Raghu2001, VanBuuren2006, VanBuuren2018, White2011}. In this approach, a regression model is specified for imputing each variable with missing values, conditional on all the other variables. The variables with missing values in $X$ are then imputed sequentially, and the procedure is iterated a few times until stable results are obtained.

SRMI can be thought of as mimicking an iterative Markov chain Monte Carlo (MCMC) algorithm under a full Bayesian joint model with flat priors, where missing values are viewed as parameters and are drawn from corresponding posterior distributions. The posterior distribution for imputing each variable is the conditional distribution of that variable given all the others, which is analogous to the SRMI approach. The SRMI standard practice for sequentially imputing variables in $X$ conditional on the other $X$ variables can be extended to also condition on response indicators, $R_1$ and $R_2$. As a generalization of SRMI under MNAR missingness, some researchers propose including missingness indicators $R_1$ and $R_2$ as predictors in regression models used for imputation \cite{Tompsett2018, Mercaldo2020}. When imputing missing values of $X_1$, for example, we might include $R_2$ as a covariate in the model that is used for imputation. $R_1$ may also be incorporated into the imputation of $X_1$ through a corresponding fixed parameter, $\delta$, used in sensitivity analysis to control the degree of MNAR dependence between $X_1$ and $R_1$. However, it is unclear how well these strategies approximate the true posterior  distribution and in what settings this approach is justified. 

In this paper, we primarily explore a particular missingness scenario where missingness in each covariate is MNAR dependent on \textit{other} variables that themselves have missing values. In this setting, we derive regression model approximations for imputing normally-distributed, binary, and categorical variables within the SRMI algorithm under this form of MNAR. This work provides theoretical justification for existing modifications of the SRMI procedure under MNAR and suggests several new extensions that may outperform existing SRMI strategies in certain settings. The paper is organized as follows: we first propose extensions of SRMI for handling MNAR missingness, including an exact imputation strategy and several simple approximations. We then compare the performance of these different approximation strategies in terms of bias in estimating downstream regression model parameters in a simulation study. We then apply these methods to handle informative missingness in a motivating study of the prevalence of BRCA1 and BRCA2 pathogenic variants among women newly-diagnosed with breast cancer, where  missingness in the BRCA1/2 status is likely related to familial history of breast cancer diagnosis, which is also only partially observed. Finally, we present a discussion.

\section{Sequential regression multiple imputation under MNAR} \label{methods}

\subsection{Deriving the conditional imputation distribution}
Assume we have a dataset consisting of $n$ independent observations in $p$ variables, denoted $X_{1},\ldots, X_{p}$. For each subject, let $R_j=0$ if $X_j$ is missing and $R_j=1$ if $X_j$ is observed. Let $X_{(-j)}$ denote the $p-1$ variables in $X$ left after excluding $X_j$, and let $R_{(-j)}$ denote the $p-1$ variables in $R$ left after excluding $R_j$. To avoid the situation where an observation has missing values for all $X$'s, we will assume that at least one of the $X$'s has no missing values for every subject. We will also assume a non-monotone pattern of missingness, by which we mean there is no $(j,k)$ pair of variables for which $R_j=0$ implies $R_k=0$. Our target of interest is some aspect of the joint distribution of $X_{1},\ldots, X_{p}$, such as the coefficients in the regression model of $X_1$ on all the other $X$'s or the mean of $X_1$. 

We propose using a sequential regression multiple imputation (SRMI) scheme to obtain $B$ complete datasets with the the missing $X$'s filled in. We then follow the standard approach \cite{Rubin1987} of analyzing each imputed dataset separately with the desired model and then combining those results to give final estimates and confidence intervals. We want to impute each variable $X_j$ with missing values from its assumed distribution given $X_{(-j)}$ and $R$, denoted $f(X_j|X_{(-j)},R)$. Some form of regression model can be used to approximate this distribution, where each regression model is tailored to the variable type for $X_j$, e.g. logistic regression if $X_j$ is binary, linear regression if $X_j$ is continuous, etc. In practice, these regression models are usually specified to have a linear combination of the variables on the right hand side, but these models could also be more flexible and include non-linear and interaction terms. The question then becomes how $R$ should be incorporated into the imputation regression models. One strategy is to include $R_{(-j)}$ directly as additional predictors in the imputation model. Since we cannot use the observed data to reliably estimate the association between $X_j$ and $R_j$, $R_j$ can be indirectly incorporated into the imputation regression model through a fixed offset term $\delta_j R_j$, where $\delta_j$ is treated as a sensitivity analysis parameter. Mercaldo et al (2020) \cite{Mercaldo2020} called this strategy multiple imputation with missing indicators (MIMI), and Tompsett et al (2018) \cite{Tompsett2018} also advocates for its general use. We will call this general strategy ``sequential regression multiple imputation with missing indicators", denoted SRMI-MI, and we focus on the particular setting where $\delta_j = 0$ under \textbf{Assumptions 1} and \textbf{2} below.

One justification for including the extra terms $R_{(-j)}$ in the imputation models is simply as a way to make the imputation model more flexible and allow the whole imputation procedure to be less reliant on the possibly restrictive assumptions of imputation models with small numbers of parameters. A more formal justification can be obtained by considering a Bayesian MCMC approach for the problem. Mimicking the ideas developed for other models \cite{Bartlett2015a,Beesley2016}, we obtain the form of the ideal conditional distribution expressed such that the imputation distribution is congenial with, or at least approximately congenial with, the desired target model of the analyst \cite{Meng1994}. Suppose that the desired target analysis model is some function of the joint distribution of $X_1,\ldots,X_p$, written as $f(X_1,\ldots,X_p)$. This joint distribution would then determine the form of any submodel based on $X$, such as the marginal distribution of $X_j$ or the conditional distribution of $X_j \vert X_{(-j)}$. Treating $(R_1,\ldots,R_p)$ as random variables, we write the joint distribution of $X_1,\ldots,X_p,R_1,\ldots,R_p$ as $f(X_1,\ldots,X_p,R_1,\ldots,R_p)$, which can be factored in a selection model form as 
\begin{equation*}
f(X_1,\ldots,X_p) \times f(R_1,\ldots,R_p|X_1,\ldots,X_p).
\end{equation*}
In the MCMC algorithm, we would ideally draw missing values of $X_j$ from the following conditional distribution: 
\begin{equation}\label{condEqRj}
f(X_j \vert X_{(-j)}, R) \propto f(X_j \vert X_{(-j)}) f(R_j \vert X, R_{(-j)}) f(R_{(-j)} \vert X)
\end{equation}
viewed as a function of $X_j$. In this expression, the distribution $f(R_j \vert X, R_{(-j)})$ is not identified using the observed data, and $f(R_{(-j)} \vert X)$ may take a complicated form in general. In order to focus our attention on a more tractable missing data setting, we make the following two assumptions:\\
\begin{rightbrace}
Assumption 1.	The $R_j$'s are conditionally independent given $X_1,\ldots,X_p$.
\newline
Assumption 2.	The missingness in $X_j$ does not depend on $X_j$, i.e. $ f(R_j|X) = f(R_j|X_{(-j)})$.
\end{rightbrace}
 The second assumption allows the missingness of one variable $R_j$ to depend on another variable $X_k, k\neq j$, which itself may be missing. In this sense, this setting is a relaxation of the usual missing at random assumption, where missingness may depend only on variables that are fully-observed given the observed data. We view the first assumption as a mild one, and it could be relaxed to have blocks of $R_j$'s be conditionally independent. The second assumption is a stronger one, and its reasonableness will depend on context of the missing data problem. Under these assumptions, we can simplify \ref{condEqRj} as follows:
\begin{equation}\label{condEqR}
f(X_j \vert X_{(-j)}, R) \propto f(X_j \vert X_{(-j)}) \prod_{k \neq j} f(R_k \vert X_j, X_{(-j)}).
\end{equation}
We see immediately that $R_j$ does not occur in this expression. Additionally, any missingness indicator $R_k$ such that $R_k \perp X_j \vert X_{(-j)}$ can also be ignored. The imputation distribution of $X_j$, therefore, will depend on $X_{(-j)}$ and any indicator $R_k, k \neq j$ such that $R_k \not\perp X_j \vert X_{(-j)}$. The distribution in \ref{condEqR} will generally be a messy expression. We can apply importance sampling methods, rejection sampling, weighting, Metropolis-Hastings algorithms, or grid-based sampling to draw directly from \ref{condEqR}. In the case of rejection sampling, for example, we could draw candidate imputations from $f(X_j \vert X_{(-j)})$ and accept the first candidate draw that satisfies $U < \prod_{k \neq j} f(R_k \vert X_j, X_{(-j)})$, where random variable $U$ is drawn from a uniform(0,1) distribution and the missingness models densities are evaluated at draws of the corresponding model parameters (see \textbf{Supplemental Material} for details). When $X_j$ is categorical, the exact form of the probability mass function can be worked out based on \ref{condEqR} as in \ref{eq:binary}. In general, we may not want to specify parametric models for the missingness probabilities, or we may prefer to impute using regression model structures. In the remainder of this paper, we will consider approximations to \ref{condEqR} that could be more easily implemented in a SRMI-MI algorithm.

A subtle but noteworthy issue is that the distribution in \ref{condEqR} does not condition on model parameters and is instead only a function of the data. For multiple imputation, drawing from the conditional distribution without parameters is usually achieved in two stages, first by drawing parameters of the model then imputing the variable from  the conditional distribution based on  that parameter value. The same technique would be used for \ref{condEqR}, in which parameters for the component distributions are \textit{drawn} from distributions that are derived using the available data. This is often implemented by fitting the corresponding component model on a bootstrap sample of the data or by making a multivariate normal approximation \cite{Little2002}. The question then becomes which subset of the data should be used to derive the distribution from which to perform these parameter draws. In a Bayesian MCMC algorithm, parameters are drawn conditional on the most recently-drawn values for all other parameters. In the missing data setting, this would suggest drawing model parameters using the most recently-imputed data for the \textit{entire} dataset of size $n$. In contrast, usual implementation of SRMI methods draw imputation model parameters for imputing $X_j$ using the data with $X_j$ observed, here called local complete case data, and treating the most recent imputations of $X_{(-j)}$ as if they were observed. It is feasible to adapt SRMI to make use of all $n$ observations, i.e. use current imputed values of $X_j$ and $X_{(-j)}$ in the estimation of the regression model for $X_j$ given all other variables. It is a easy to show that, theoretically, either approach can be used when missingness in $X_j$ is independent of $R_j$. In practice, imputing within SRMI based on all $n$ observations may be preferred simply because the increased sample size may give better estimates of the relationship between each $X_j$ and $R_{(-j)}$. This approach is used in our simulations and data analysis.

\subsection{Regression model approximations for imputing binary, categorical and continuous variables}
In this section, we approximate the imputation distribution proportional to \ref{condEqR} under different assumptions about the distributions of the variables in $X$. 

\subsubsection{Imputing binary variables} \label{binary}
Suppose we want to impute binary variable $X_1$ and that the distribution for $X_1 \vert X_{(-1)}$ is well-approximated by a logistic regression model as follows:
\begin{align*}
P1 &= P(X_1=1|X_{(-1)}) = \text{expit}(\theta_0+ \Sigma_{j=2}^p \theta_j X_j)
\end{align*}
where $\text{expit}(u) = \exp(u)/(1+\exp(u))$. Let $PRj(x_1)$ denote the probability of observing $X_j$ given $X_{(-j)}$ with $X_1 = x_1$. We note that $PRj(x_1)$ can be a function of all the $X$'s except $X_j$, but for conveneunce we just use the notation $PRj(x_1)$. Thus, for example, $PR2(1) = P(R_2=1|X_1=1, X_3, \hdots, X_p)$. Following this notation and accounting for proportionality, we can express \ref{condEqR} as $P(X_1=1|X_{(-1)},R) = A/(A+B)$ where 
\begin{align*}
&A= P1 \prod_{j=2}^p  PRj(1)^{R_j}\left[1-PRj(1)\right]^{1-R_j} \\
\text{and }\hspace{0.5cm}&B= (1-P1) \prod_{j=2}^p PRj(0)^{R_j}\left[1-PRj(0)\right]^{1-R_j} 
\end{align*}
This expression simplifies as follows:
\begin{align} \label{eq:binary}
\text{log}&\left[\frac{P(X_1=1 \vert X_{(-1)}, R)}{1-P(X_1=1 \vert X_{(-1)}, R)} \right] \\
& = \text{log}\left[ \frac{P1}{1-P1}\right] + \sum_{j=2}^p \left\{ R_j \text{log}\left[ \frac{PRj(1)}{PRj(0)} \right] +  (1-R_j) \text{log}\left[ \frac{1-PRj(1)}{1-PRj(0)} \right] \right\} \nonumber \\
& = \theta_0+ \sum_{j=2}^p \theta_j X_j + \sum_{j=2}^p \left\{ R_j \text{log}\left[ \frac{PRj(1)}{PRj(0)} \right] +  (1-R_j) \text{log}\left[ \frac{1-PRj(1)}{1-PRj(0)} \right] \right\} \nonumber
\end{align}
This can also be rewritten as 
\begin{align*}
\text{log}&\left[\frac{P(X_1=1 \vert X_{(-1)}, R)}{1-P(X_1=1 \vert X_{(-1)}, R)} \right] \\
& = \theta_0+ \sum_{j=2}^p \theta_j X_j + \sum_{j=2}^p  R_j \text{log}\left[ \frac{PRj(1)}{PRj(0)}\frac{\{1-PRj(0)\}}{\{1-PRj(1)\}} \right] +  \sum_{j=2}^p  \text{log}\left[ \frac{1-PRj(1)}{1-PRj(0)} \right] 
\end{align*}
We now consider several special cases and then propose a general strategy for imputation of a binary variable. \\

\noindent \textbf{\underline{Binary Special Case 1: logistic missingness with main effects}.}\\
\noindent Suppose that the model for missingness for each variable $X_j$ can be expressed as follows:
\begin{align} \label{eq:logit}
PRj(X_1) &=  P(R_j=1|X_1, X_{(-1)})  = \text{expit}( \phi_{j0} +  \Sigma_{k \neq j}  \phi_{jk} X_k).
\end{align}
In this case, \ref{eq:binary} can be simplified as
\begin{align*}
&\text{logit}[ P(X_1=1 \vert X_{(-1)}, R)] = \theta_0+ \sum_{j=2}^p \theta_j X_j + \sum_{j=2}^p \phi_{j1} R_j \nonumber \\
& +   \sum_{j=2}^p \text{log}[ 1+\exp(\phi_{j0}  + \sum_{k=2, k \neq j}^p \phi_{jk} X_k ) ] -\text{log}[ 1+\exp(\phi_{j0} +  \phi_{j1}  + \sum_{k=2, k \neq j}^p \phi_{jk} X_k ) ] 
\end{align*}
In the special case where $p=3$ and $X_2$ and $X_3$ are binary, all the terms involving the $log$'s can be simplified and combined with $\theta_0$ and the $\theta_j X_j$'s, and the final expression is simply a linear combination of $X_2, \hdots, X_p$ and $R_2, \hdots, R_p$ as follows:
\begin{equation} \label{eq:1}
\text{logit}\left[P(X_1=1|X_{(-1)}, R)  \right] = \omega_0 +\sum_{j=2}^p \omega_j X_j + \sum_{j=2}^p \omega_{Rj} R_j .
\end{equation}
In general for $p>3$ and for non-binary $X_2$ and $X_3$, \ref{eq:binary} does not reduce to this simple additive form. However, a first order Taylor series approximation of the logarithm terms (assuming all values of $\phi_{jk}, k \ne j$ are small) does lead to \ref{eq:1} as an approximation to the desired imputation distribution. A second order Taylor series approximation results in the following regression model structure:
\begin{align}\label{eq:3b}
\text{logit}&\left[P(X_1=1|X_{(-1)},R)\right] \approx \alpha_0 + \sum_{k=2}^p \alpha_k X_k + \sum_{k=2}^p \alpha_{Rk} R_k +  \sum_{j=2}^p \sum_{k=2}^p  \alpha_{2jk}  X_j  X_k 
\end{align}
 to impute $X_1$, i.e. including interactions between the $X$'s. \\

\noindent \textbf{\underline{Binary Special Case 2: interactions in logistic missingness model}.} \\
\noindent Suppose instead that the missingness models include interactions between other covariates and $X_1$. For simplicity, we will assume $p=3$. Suppose that
\begin{align*}
\text{logit}\left[PR2(X_1)\right] = \phi_{20} +  \phi_{21} X_1 + \phi_{23} X_3 +\phi_{24} X_1 X_3\\
\text{logit}\left[PR3(X_1)\right] =  \phi_{30} +  \phi_{31} X_1 + \phi_{32} X_2 + \phi_{34}X_1 X_2
\end{align*}
In this case, the imputation takes the following form:
\begin{align*}
\text{logit}&\left[P(X_1=1|X_2,X_3,R_2,R_3)\right] =\theta_0 +\theta_2 X_2 + \theta_3 X_3 \\
& + R_2 \left[\phi_{21} + \phi_{24}X_3\right] + R_3 \left[\phi_{31} + \phi_{34}X_2\right] \\
& +  \text{log}\left[1+\exp(\phi_{20} + \phi_{23} X_3 )\right]-\text{log}\left[ 1+\exp(\phi_{20} +  \phi_{21}  + \phi_{23} X_3 +\phi_{24}  X_3) \right]    \\
&+  \text{log}\left[1+\exp(\phi_{30} + \phi_{32} X_2 )\right]-\text{log}\left[ 1+\exp(\phi_{30} +  \phi_{31}  + \phi_{32} X_2 +\phi_{34}  X_2) \right].
\end{align*}
 Using the same logic as before and applying a first order Taylor series approximation, we can express the imputation distribution as follows:
\begin{align}\label{eq:2}
& \text{logit}\left[P(X_1=1|X_2,X_3,R_2,R_3)  \right]\\
& \approx \omega_0 +\omega_2 X_2 + \omega_3 X_3 +\omega_{R2} R_2 + \omega_{R3} R_3 +\omega_{3,R2} X_3 R_2 + \omega_{2,R3} X_2 R_3 \nonumber
\end{align}
For $p>3$, we can similarly approximate the imputation distributions by including interactions between the $X$'s and missingness indicators.\\

\noindent \textbf{\underline{Binary General Case}.} \\
Suppose now that variables $X_2, \hdots, X_p$ have some unspecified form and we allow $PRj(X_1) = P(R_j = 1 \vert X)$ to take more general (e.g. non-logistic) form. We notice that \ref{eq:binary} resembles a logistic regression model with predictors $X_{(-1)}$ and a term that is a function of the missingness indicators, $R_{(-j)}$, and the probabilities of missingness, $PRj(X_1)$. Guided by \ref{eq:binary}, we propose the following strategy for imputing missing values of $X_1$ within each iteration of a chained equations imputation algorithm:
\begin{enumerate} 
\item  For each $j>1$, fit a model (e.g. logistic or probit regression or even a regression tree) to the current imputed dataset of size $n$ for the probability that $X_j$ is observed. 
\item For each observation and each $j>1$, use these model estimates to calculate the probability that $X_j$ is observed with $X_1$ set to 0 and with $X_1$ set to 1 to give $PRj(0)$ and $PRj(1)$, respectively. To calculate these probabilities, use the most recent imputed values for $X_{(-j)}$.
\item Define new variables 
\begin{equation}\label{offsetbinary1}
Z_j=R_j \log\left[\frac{PRj(1)}{PRj(0)}\right] + (1-R_j) \log\left[\frac{1-PRj(1)}{1-PRj(0)}\right].
\end{equation}
\item Impute $X_1$ using the following model:
\begin{equation} \label{offsetbinary2}
\hspace{-0.5cm}\text{logit}\left[P(X_1=1|X_{(-1)},R_2,R_3,Z_2,Z_3)\right] = \omega_0 + \sum_{k=2}^p \omega_k X_k + \sum_{k=2}^p Z_j
\end{equation}
 where the $\omega$'s are first drawn from an approximation to their posterior distribution derived from a model fit to the full imputed dataset and where $\sum_{k=2}^p Z_j$ is a fixed offset (with coefficient equal to 1). 
\end{enumerate}

\subsubsection{Imputing multinomial variables} \label{multinomial}
Now, we suppose that $X_1$ is a categorical variable taking values in $0,1,\hdots, S$ and that the distribution for $X_1 \vert X_{(-1)}$ is well-approximated by a multinomial regression as follows:
\begin{align*}
PS &= P(X_1=s|X_{(-1)}) = \frac{\exp(\theta_{0s} + \sum_{j=2}^p\theta_{js} X_j )}{1+\sum_{r=1}^{S}\exp(\theta_{0r} + \sum_{j=2}^p\theta_{jr} X_j )}
\end{align*}
where all $\theta_{j0}$'s are equal to zero. As in the derivation of \ref{eq:binary}, we can write the imputation distribution as follows:
\begin{align} \label{eq:multinomial}
& \text{log}\left[\frac{P(X_1=s \vert X_{(-1)}, R)}{P(X_1=0 \vert X_{(-1)}, R)} \right] \\
& = \theta_{0s} + \sum_{j=2}^p \theta_{js} X_j + \sum_{j=2}^p \left\{ R_j \text{log}\left[ \frac{PRj(s)}{PRj(0)} \right] +  (1-R_j) \text{log}\left[ \frac{1-PRj(s)}{1-PRj(0)} \right] \right\} \nonumber
\end{align}
where $PRj(s)$ corresponds to the probability of observing $X_j$ with $X_1 = s$.\\
\indent In the special case where $PRj(X_1)$ corresponds to a logistic regression with main effects such that
\begin{align*}
\text{logit}(P(R_j=1|X_1 = s, X_{(-1)})) = \phi^s_{0j} + \sum_{k=2, k\neq j}^p \phi^s_{kj} X_k,
\end{align*}
we have the following for $s>0$:
\begin{align} \label{eq:multinomial1}
&\text{log}\left[\frac{P(X_1=s \vert X_{(-1)}, R)}{P(X_1=0 \vert X_{(-1)}, R)} \right] = \theta_{0s} + \sum_{j=2}^p \theta_{js} X_j\\
& + \sum_{j=2}^p \left\{  R_j [\phi^s_{0j} + \sum_{k=2, k\neq j}^p \phi^s_{kj} X_k] +  \text{log}[ 1+\exp(\phi^s_{j0}  + \sum_{k=2, k \neq j}^p \phi^s_{jk} X_k ) ] \right\} \nonumber \\
& - \sum_{j=2}^p \left\{  R_j [\phi^0_{0j} + \sum_{k=2, k\neq j}^p \phi^0_{kj} X_k] +  \text{log}[ 1+\exp(\phi^0_{j0}  + \sum_{k=2, k \neq j}^p \phi^0_{jk} X_k ) ] \right\} \nonumber
\end{align}
A first order Taylor series approximation of \ref{eq:multinomial1} suggests a regression of the form:
\begin{equation} \label{eq:multiapprox}
\text{log}\left[\frac{P(X_1=s \vert X_{(-1)}, R)}{P(X_1=0 \vert X_{(-1)}, R)} \right]\approx \omega_{0s} +\sum_{j=2}^p \omega_{js} X_j + \sum_{j=2}^p \omega^s_{Rj} R_j + \sum_{j=2}^p \sum_{k=2, k \neq j}^p \omega^s_{RjXk} R_j X_k.
\end{equation}
In other words, we can include the missingness indicators and their interactions with $X$ as additional predictors. If we can further assume no interaction between $X_1$ and the other $X$'s in the model for the missingness of $X_j$, then $\phi^s_{kj}$ takes a single value across $s=1,\ldots,S$ for $k=2,\ldots,p$, $k \neq j$. In this case, we have 
\begin{equation} \label{eq:multiapprox2}
\text{log}\left[\frac{P(X_1=s \vert X_{(-1)}, R)}{P(X_1=0 \vert X_{(-1)}, R)} \right]\approx \alpha_{0s} +\sum_{j=2}^p \alpha_{js} X_j + \sum_{j=2}^p \alpha^s_{Rj} R_j,
\end{equation}
indicating that we should just include the missingness indicators in the imputation model. \\
\indent For more general missingness mechanisms, we can apply a generalization of the offset strategy of \ref{offsetbinary2} where we define offsets:
\begin{equation}\label{offsetmult1}
Z_{js} =R_j \text{log}\left[ \frac{PRj(s)}{PRj(0)} \right] +  (1-R_j) \text{log}\left[ \frac{1-PRj(s)}{1-PRj(0)} \right]  
\end{equation}
and impute from a regression model as follows:
\begin{equation} \label{offsetmult2}
\text{log}\left[\frac{P(X_1=s \vert X_{(-1)}, R)}{P(X_1=0 \vert X_{(-1)}, R)} \right] = \omega_{0s} + \sum_{k=2}^p \omega_{ks} X_k + \sum_{k=2}^p Z_{ks}
\end{equation}
where $\sum_{k=2}^p Z_{ks}$ is a fixed offset.


\newpage
\subsubsection{Imputing continuous variables}  \label{continuous}
We now suppose that $X_1$ follows some continuous distribution defined on the real line. First, we will consider the special case where $X_1$ is normally-distributed given $X_{(-1)}$. Then, we will propose a strategy for more general $X_1$.  \\

\noindent \textbf{\underline{Continuous Special Case 1: imputing normally-distributed variable}.} \\
Suppose first that $X_1$ is normally distributed such that  $X_1|X_{(-1)} \sim N(\theta_0 + \sum_{k=2}^p \theta_k X_k, \sigma^2)$. Suppose further that the probability of observing $X_j$ is given by 
\begin{equation*}
\text{logit}\left[P(R_j=1|X)\right] = \phi_{j0} + \sum_{k\neq j} \phi_{jk} X_k.
\end{equation*}
\noindent Following \ref{condEqR}, we can express the imputation model for $X_1$ as 
\begin{align} \label{eq:4}
 &f(X_1 \vert X_{(-1)}, R_{(-1)}) \propto f(X_1 \vert X_2, \hdots, X_p) \prod_{k=2}^p f(R_k \vert X_1, \hdots, X_p)\\
&\propto \exp\left(-\frac{[X_1 -(\theta_0+ \sum_{k=2}^p \theta_k X_k)]^2}{2 \sigma^2}\right) \times  \prod_{k=2}^p \frac{\exp(R_k[\phi_{k0} +  \sum_{s \neq k}  \phi_{ks} X_s ]) }{1+\exp(\phi_{k0} +  \sum_{s \neq k}  \phi_{ks} X_s)}  \nonumber
\end{align}
 Consider the special case where $p=3$ (so $X = (X_1, X_2, X_3)$). The two terms in \ref{eq:4} are respectively a bell-shaped curve and the product of two separate bounded sigmoid functions as a function of $X_1$. The sigmoid curve for $f(R_2|X)$ will be increasing in $X_1$ for one value of $R_2$ and decreasing for the other value, and likewise $f(R_3|X)$ will be increasing in $X_1$ for one value of $R_3$ and decreasing for the other value. To represent a valid distribution, the product in \ref{eq:4} has to be normalized to integrate to 1. More generally, it is clear that the full conditional distribution of $X_1$ will depend on $R_k$, assuming $\phi_{k1} \ne 0$. Additionally, the conditional distribution of $X_1$ is \textit{not} symmetric and its mean is no longer given by $\theta_0+\sum_{k=2}^p \theta_k X_k$. While it is feasible to draw from the distribution proportional to \ref{eq:4} exactly, we will explore approximations that may be easier to draw from in practice.\\

\noindent \textbf{Approximation Strategy 1:} An intuitive approximation of \ref{eq:4} would be to draw $X_1$ from the following normal distribution:
\begin{equation} \label{eq:4b}
N(\omega_0 + \sum_{k=2}^p \omega_k X_k + \sum_{k=2}^p \omega_{Rk} R_k, \tau^2).
\end{equation}
This strategy can be justified as a second order Taylor series approximation of \ref{eq:4} as follows. Assuming $\phi_{jk}$ is small for all $k$,
\begin{align*}
& \log\left[f(R_j|X)\right] \approx R_j[\phi_{j0} + \sum_{k \neq  j} \phi_{jk} X_k]+ \log(1+\exp(\phi_{j0})) \\
& + \frac{\exp(\phi_{j0})}{1+\exp(\phi_{j0})} [\phi_{j1} X_1, \hdots, \phi_{jp} X_p]^T +
\frac{\exp(\phi_{j0} )}{\left[1+\exp(\phi_{j0})\right]^2} [\phi_{j1} X_1, \hdots, \phi_{jp} X_p]^{\otimes 2}
\end{align*} 
where $\phi_{jj}=0$. Combining these expressions with the form for $\log(f(X_1|X_{(-1)}))$ and collecting terms multiplied by $X_1$ in \ref{eq:4}, we obtain a linear regression in the form of \ref{eq:4b}.\\
\indent If the association between the $X$'s and the $R$'s is stronger, then this Taylor series approximation may be less accurate, and a more involved approach to drawing values of missing $X$'s is needed. For example, we notice from equation \ref{eq:4} that $X_2$ appears in $f(X_1|X_{(-1)})$ and may also be included in the various missingness models, suggesting something more general than a linear term in $X_2$ may be needed for imputing $X_1$. We propose including a spline function of $X_2$. Similar spline terms could also be included for other covariates in the imputation model. This results in the following approximate imputation distribution:
\begin{align} \label{eq:4c}
N(s_2(X_2) + s_3(X_3) + \hdots + s_p(X_p) + \sum_{k=2}^p \omega_{Rk} R_k , \tau^2).
\end{align} 
where $s_k(X_k)$ denotes a spline function of covariate $X_k$.
 The presence of the product of sigmoid curves in \ref{eq:4} modifies both the spread and skewness of the imputation distribution. We will ignore the skewness, but we could accommodate the spread by letting it depend on the values of $R$. Thus, another level of approximation would be drawing $X_1$ from a normal distribution 
\begin{equation}\label{eq:4d}
N(s_2(X_2) + s_3(X_3) + \hdots + s_p(X_p)+ \sum_{k=2}^p \omega_{Rk} R_k , \tau^2_{(R2,R3,\hdots,Rp)}).
\end{equation} 
As an even more flexible approximation, we might allow the variance to depend on $R$ and incorporate interactions between $X$ and $R$ in the mean structure of the imputation distribution. The approximations in \ref{eq:4c} and \ref{eq:4d} could be incorporated into a sequential regression multiple imputation procedure, provided the software being used had the ability to include splines instead of simple linear terms in the mean structure of the regression model. In practice, a large value of $n$ may be required to actually fit the largest of the above models during the imputation procedure. To build in even more flexibility in the imputation model, we might take a generally more robust approach to multiple imputation, such as predictive mean matching \cite{Morris2014, Schenker1996} or random forests \cite{Shah2014}, conditioning on $R_2, \hdots, R_p$ in addition to other variables when imputing $X_1$. \\

\noindent \textbf{Approximation Strategy 2:} Rather than approximating the mean structure of \ref{eq:4} using Taylor series approximations, we could instead consider the mode of the distribution in \ref{eq:4}, which we call $mode(X_{(-1)},R_{(-1)})$. Assuming the distribution in \ref{eq:4} is uni-modal, then we might impute missing $X_1$ from $N(mode(X_{(-1)},R_{(-1)}), \tau^2)$. Taking the derivative with respect to $X_1$ of the log of \ref{eq:4} leads to the following expression:
\begin{align}\label{eq:4e}
-\frac{X_1 - (\theta_0+\sum_{k=2}^p \theta_k X_k )}{\sigma^2} + \sum_{k=2}^p \phi_{k1}[R_k-PRk(X_1)] 
\end{align}
where $PRk(X_1) = P(R_j=1|X) = \text{expit}\left( \phi_{j0} + \sum_{k\neq j} \phi_{jk} X_k \right)$ is viewed as a function of $X_1$. Assuming a uni-modal distribution, we can obtain the $mode(X_{(-1)},R_{(-1)})$ by setting \ref{eq:4e} equal to 0 and solving for $X_1$. Finding this mode is numerically feasible, but the form of \ref{eq:4e} suggests an alternative approach for imputing $X_1$ within the iterative imputation algorithm:
\begin{enumerate}
\item  For each $j>1$, fit a logistic regression model to the current imputed dataset of size $n$ for the probability that $X_j$ is observed. 
\item Using the most recent imputed values for $X_1$ and the latest estimates of $\phi$ and $PRj(X_1)$ obtained in step 1, define new variables $Z_j= \phi_{j1}[R_j - PRj(X_1)]$ for each $j>1$.
\item Impute $X_1$ using the following model:
\begin{equation}\label{offsetnormal}
N(\omega_0 + \sum_{k=2}^p \omega_k X_k  +\sigma^2 \sum_{k=2}^p Z_k, \tau^2)
\end{equation}
where the $\omega$'s are drawn from the approximation to their posterior distribution obtained by fitting \ref{offsetnormal} to the full imputed dataset and $\sigma^2 \sum_{k=2}^p Z_k$ is treated as an offset using the estimate of $\sigma^2$ obtained from fitting the model for $X_1$ given $X_2, \hdots, X_p$ to the complete data. Alternatively $\sum_{k=2}^p Z_k$ could be added as another predictor in the imputation model.
\end{enumerate}

\noindent \textbf{\underline{Continuous General Case: non-normal continuous variable}.} \\
Suppose $X_1$ takes a more general non-Gaussian continuous form and that $X_2 \ldots X_p$ take unspecified forms. For this case we may first transform $X_1$ so that the conditional distribution of $X_1|X_2,\ldots,X_p$ is approximately Gaussian with constant variance $\sigma^2$. Using the intuition developed for normally-distributed $X_1$ considered above, we propose the following three strategies for approximating the conditional imputation distribution for $X_1$ in \ref{condEqR} using one of the following three imputation distributions: 
\begin{equation}\label{eq:5}
N(\omega_0 + \sum_{k=2}^p \omega_k X_k + \sum_{k=2}^p \omega_{Rk} R_k, \tau^2),
\end{equation} 
\begin{equation}\label{eq:5b}
N(\sum_{k=2}^p s_k(X_k) + \sum_{k=2}^p \omega_{Rk} R_k, \tau^2),
\end{equation} 
where $s_k(X_k)$ is a spline function of $X_k$, and 
\begin{equation}\label{eq:5c}
N(\omega_0 + \sum_{k=2}^p \omega_k X_k + \sigma^2 \sum_{k=2}^p Z_k, \tau^2),
\end{equation} 
where $Z_k = \phi_{k1}[R_k - PRk(X_1)]$ is a constructed variable based on estimated probability of observing $X_k$, $PRk(X_1) = P(R_k = 1 \vert X_{(-k)})$, obtained using the most recent imputed data.

\section{Simulation Studies}  \label{simulations}

\subsection{Simulation Set-up}
We performed numerical studies to investigate the performance of the proposed method under different missingness and X distribution settings. For each setting, we generate 200 simulated datasets with 2000 subjects each. In each simulated dataset, we generate 5 correlated variables under two different scenarios. In the first scenario, we simulated 5 multivariate normal variables $X_1, \hdots, X_5$ with mean 0, unit variances, and covariances $\Sigma_{jk} = cov(X_j, X_k)$ as follows: $\Sigma_{12} = 0.4$, $\Sigma_{14}$ = $\Sigma_{35}$ = 0.3, $\Sigma_{13}$ = $\Sigma_{25}$ = $\Sigma_{34}$ = 0.2, and all remaining covariances equal to 0.1. In the second scenario, covariates $X_1, X_2$, and $X_3$ are dichotomized to take the value 1 if the drawn value is above zero. We then impose roughly 25-50\% missingness in each of $X_1$, $X_2$, and $X_3$ under the following models:
\begin{align*}
\text{logit}(P(R_1 = 1 \vert X_2, X_3, X_4, X_5) = \phi X_2 + \phi X_3 + \rho X_4 + \rho X_5 \\
\text{logit}(P(R_2 = 1 \vert X_1, X_3, X_4, X_5) = \phi X_1 + \phi X_3 + \rho X_4 + \rho X_5 \\
\text{logit}(P(R_3 = 1 \vert X_1, X_2, X_4, X_5) = \phi X_1 + \phi X_2 + \rho X_4 + \rho X_5 
\end{align*}
where $\phi$=0, 0.25, 0.50, 0.75, 1, or 1.5 and $\rho$ was either 0 or 1. Corresponding complete case probabilities ranged between 12\% and 50\%. \\
\indent For each simulated dataset in each setting, we obtained 10 multiple imputations for missing values in $X_1$, $X_2$, and $X_3$ using the following methods:
\begin{enumerate}
    \item SRMI: usual chained equations assuming missing at random
    \item SRMI-MI: method SRMI + adjusting for missingness indicators as in \ref{eq:1} and \ref{eq:4b}
    \item SRMI-Interactions R: method SRMI-MI + adjusting for missingness indicator-covariate interactions as in \ref{eq:2} and \ref{eq:4d}
    \item SRMI-Interactions X: method SRMI-MI + adjusting for missingness covariate-covariate interactions as in \ref{eq:3b}
    \item SRMI-TriCube: adjusting for missingness indicators and cubic splines for other covariates as in \ref{eq:4c}
    \item SRMI-Offset(Normal): method SRMI + estimated offset as in \ref{offsetnormal}
    \item SRMI-Offset(Binary): method SRMI + estimated offset as in \ref{offsetbinary2}
    \item SRMI-Exact: imputing from ``exact" distribution proportional to \ref{condEqR}, using drawn missingness model parameters
\end{enumerate}
The SRMI-Exact method imputes missing values from the correct conditional distribution after estimating missingness model parameters in the observed data. This method serves as a benchmark for the various (more easily implemented) approximations considered. 

For scenarios with normally-distributed or binary $X_1$, $X_2$, and $X_3$, we performed imputation using a subset of the above methods relevant for the corresponding covariate distributions as motivated by our derivations above. For the SRMI-Offset(Normal), SRMI-Offset(Binary), and SRMI-Exact methods, we assumed a logistic regression model structure for missingness in each variable, and we estimate or draw corresponding missingness model parameters using the most recently imputed data. Parameters in the missingness model can be estimated well, as demonstrated by simulation \textbf{Figure A.1} in the \textbf{Supplemental Material}. For each simulation setting and imputation strategy combination, we obtained point estimates  for (1) the mean of $X_1$ and for (2) regression coefficients from a model for $X_1 \vert X_2, X_3, X_4, X_5$ using the multiply imputed data and Rubin's combining rules. We then calculated the average bias, empirical variance of the point estimates, and the coverage rate of 95\% confidence intervals across the 200 simulated datasets. \\

\subsection{Simulation Results}
\begin{figure}[htbp!]
  \centering
\caption{Bias for mean of $X_1$ across 200 simulated datasets after applying various imputation strategies$^1$}
\subfloat[][Normally-distributed $X_1$, $X_2$, and $X_3$]{\includegraphics[trim={0cm 0cm 0cm 0.8cm}, clip, width=5.5in]{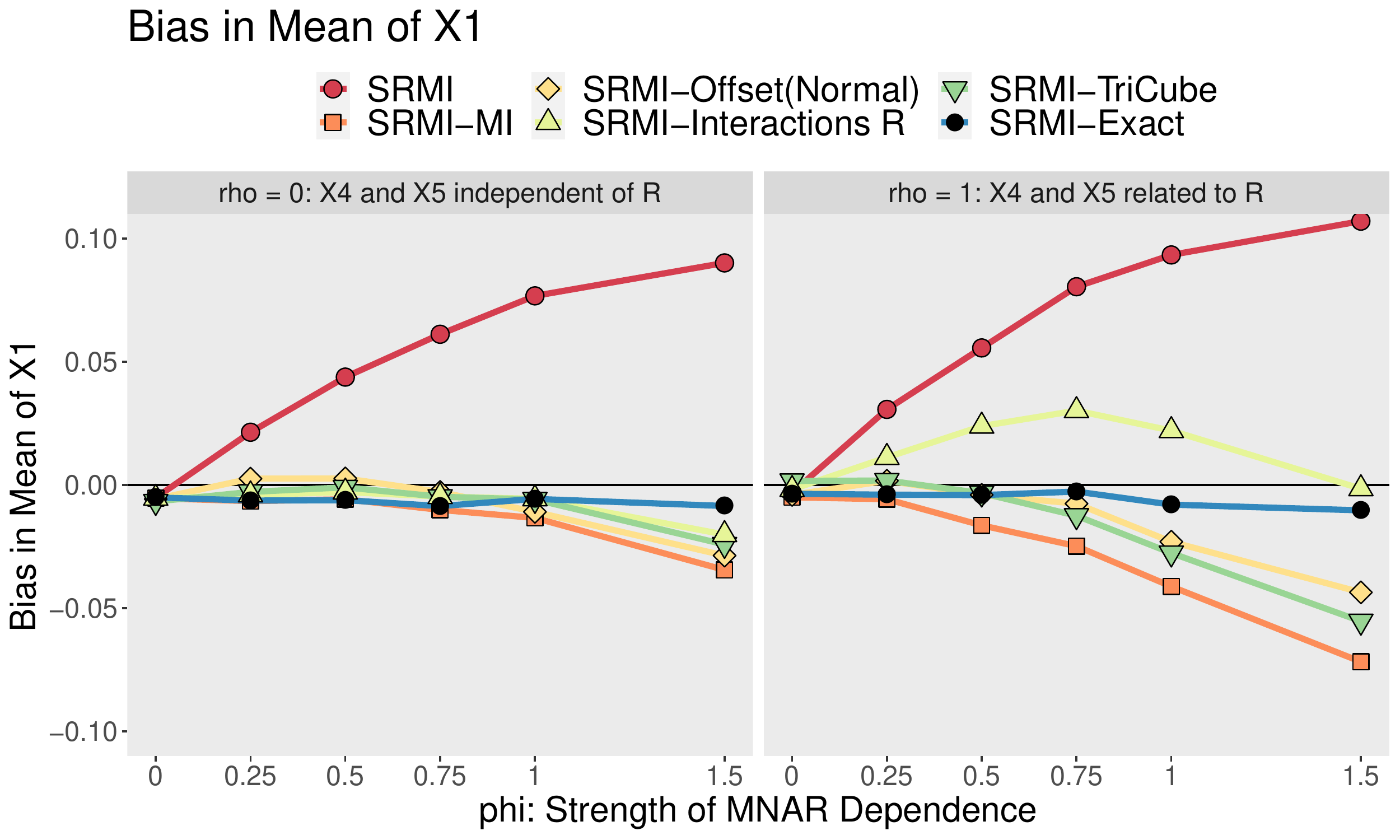}}\\
\subfloat[][Binary $X_1$, $X_2$, and $X_3$]{\includegraphics[trim={0cm 0cm 0cm 0.8cm}, clip, width=5.5in]{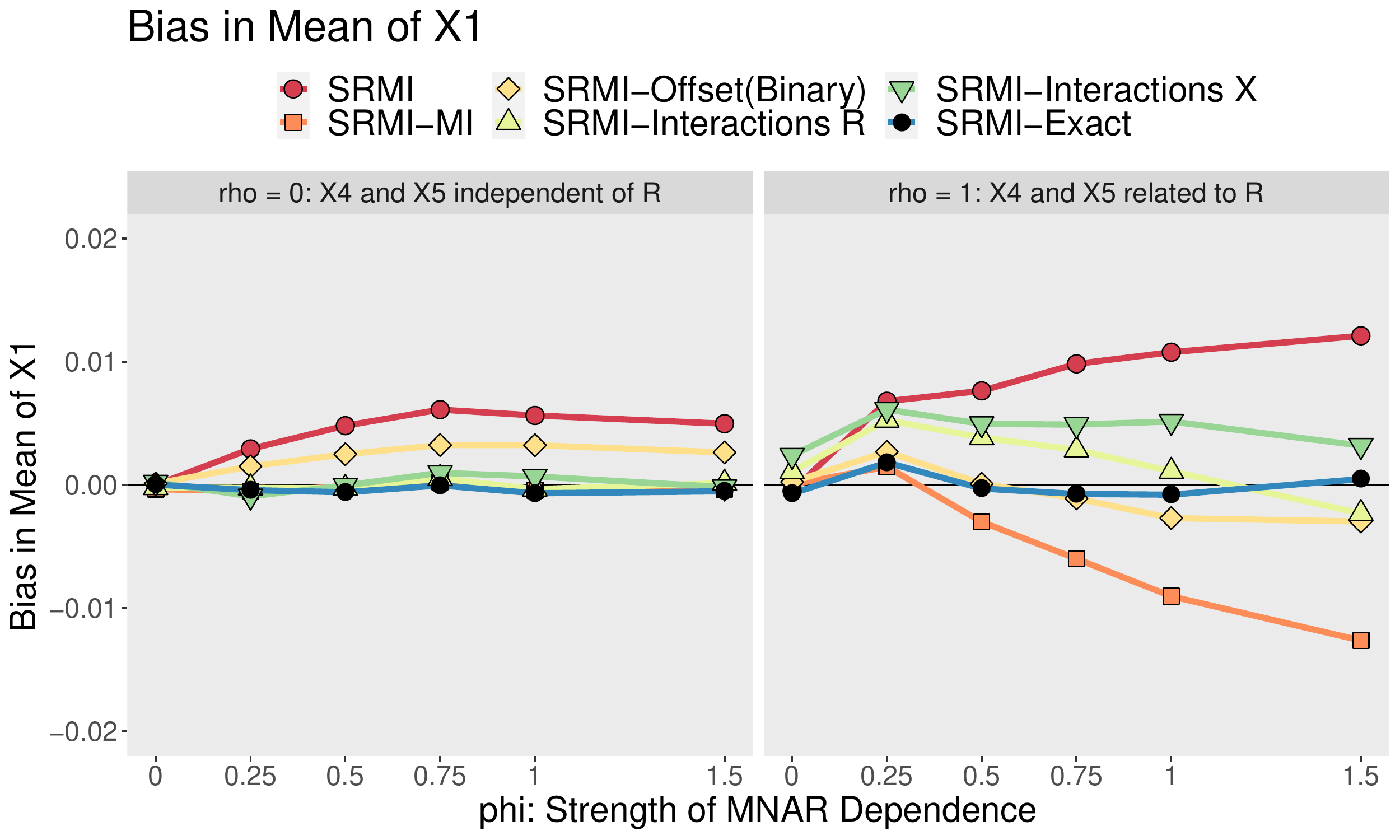}}
\caption*{ \footnotesize $^1$ Results shown for $M=10$ imputed datasets.  }
\label{mean_bias}
\end{figure}

\begin{figure}[htbp!]
  \centering
\caption{Relative variance for estimated mean of $X_1$ across 200 simulated datasets after applying various imputation strategies, relative to analysis of the full data with no missingness$^1$}
\subfloat[][Normally-distributed $X_1$, $X_2$, and $X_3$]{\includegraphics[trim={0cm 0cm 0cm 0.8cm}, clip, width=5.5in]{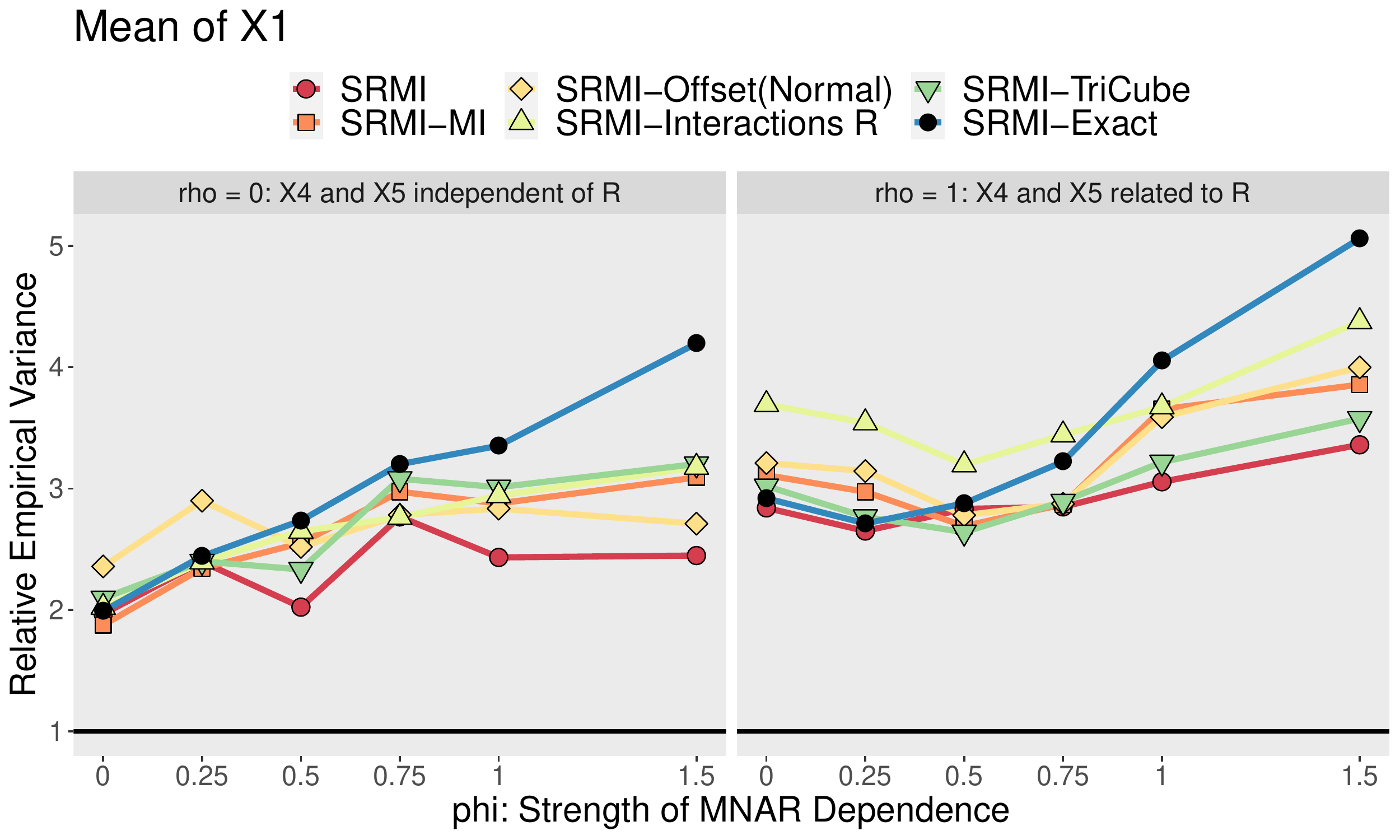}}\\
\subfloat[][Binary $X_1$, $X_2$, and $X_3$]{\includegraphics[trim={0cm 0cm 0cm 0.8cm}, clip, width=5.5in]{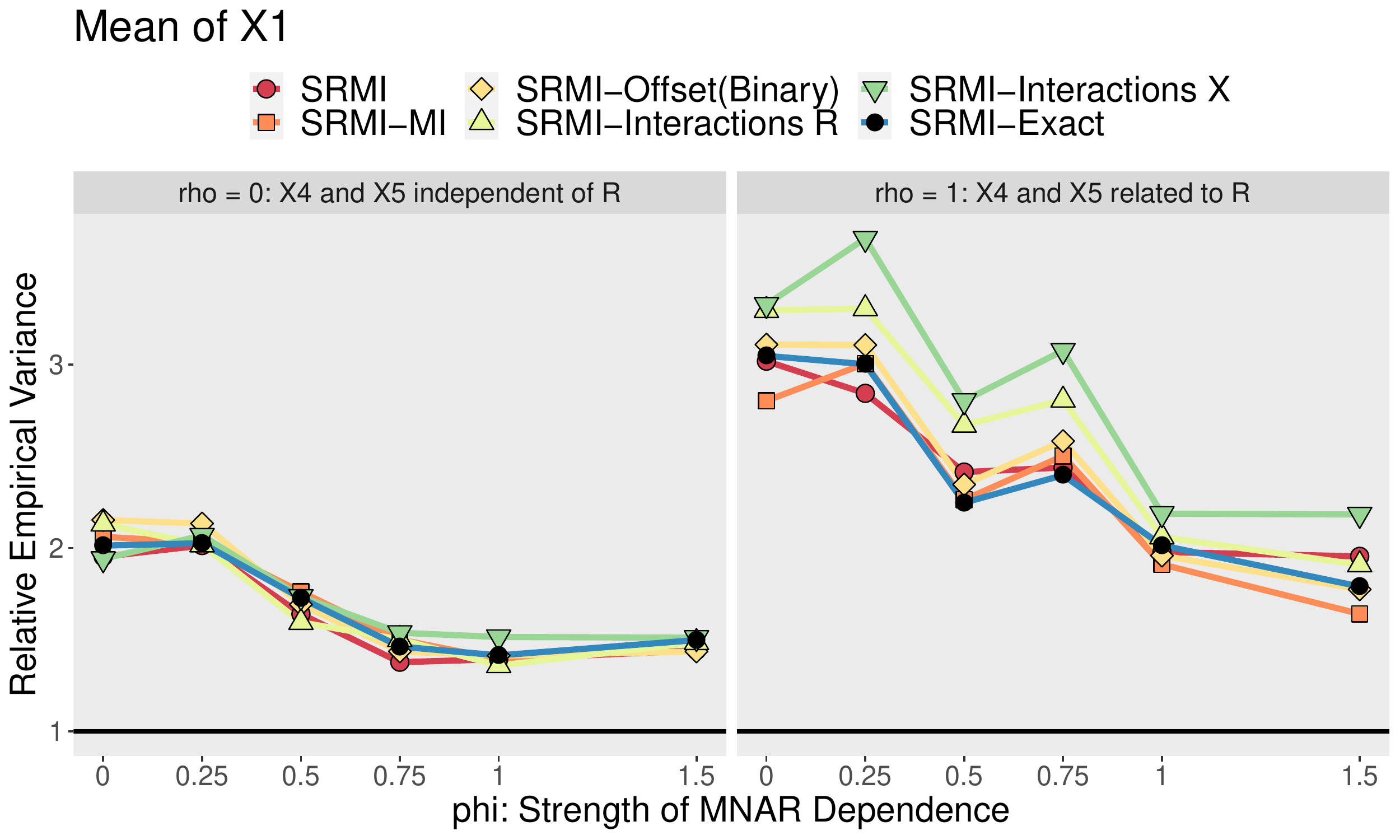}}
\caption*{ \footnotesize $^1$ Results shown for $M=10$ imputed datasets.  }
\label{mean_relvar}
\end{figure}

\indent \textbf{Figure \ref{mean_bias}} shows the bias in estimating the mean of $X_1$ for different imputation methods. Under MAR ($\phi = 0$), none of the methods gave substantial bias. For both normally-distributed and binary variables, SRMI produced substantial bias (e.g., absolute bias of 0.10 for normal $X_1$) under MNAR ($\phi \neq 0$). In both normal and binary settings, all MNAR adjustment methods considered resulted in similar or reduced bias relative to SRMI (e.g., SRMI-MI resulted in up to 80\% reduction in bias relative to SRMI for normal $X_1$). The SRMI-MI method worked well to reduce bias from MNAR missingness when (1) MNAR missingness was weak or (2) missingness did not depend on the continuous variables ($\rho$ = 0). \\
\indent In the setting with very strong MNAR missingness or missingness dependent on continuous covariates, the SRMI-MI approximation resulted in large residual bias (e.g. absolute bias of -0.07). For imputation of normally-distributed covariates, the SRMI-Exact method was the only approach that consistently produced good properties in terms of bias. Imputation models using more complicated functions of predictors (e.g. interactions, splines) often provided smaller bias relative to SRMI-MI but did not perform as well as imputation using the ``exact" conditional distribution, particularly when $\rho \neq 0$. For imputation of binary covariates, the offset approach generally performed well in terms of bias reduction, particularly when missingness model parameters were fixed to the simulation truth (not shown). Some small residual bias was seen for the offset method when missingness model parameters were estimated. Although not shown, complete case analysis resulted in very large bias in all simulation settings considered. Biases for regression model coefficients are presented in \textbf{Figure A.3}. Results are similar.\\
\indent \textbf{Figure \ref{mean_relvar}} shows the empirical variance of point estimates for the mean of $X_1$, relative to analysis of the full data with no missingness. Under MAR, there is at most a small increase in the variability due to the extensions of the standard SRMI method. Inclusion of additional interaction terms (between missingness indicators and covariates or between covariates themselves) in the imputation models resulted in larger empirical variances. In the setting with normally-distributed covariates, SRMI-Exact imputation resulted in larger empirical variance when the MNAR missingness was very strong. However, coverage rates (\textbf{Figure A.2}) were similar to other methods, indicating that wider confidence intervals may be a necessary trade-off for little bias.\\
\indent We also performed some additional simulations in the setting where $X_1$-$X_3$ are binary variables with very low prevalences (1\%). In this particular setting, additional adaptations are needed to ensure good estimation of the prevalence of $X_1$ in the imputed data. Results are presented in \textbf{Supplemental Material}.

\section{Prevalence of genetic pathogenic variants in breast cancer patients}  \label{data}
The methodological development in this paper was motivated by missing data challenges for the ICanCare study. This study consists of women aged 20 to 79 who were newly diagnosed with breast cancer between July 2013 and August 2015 and are part of the Surveillance, Epidemiology, and End Results (SEER) registries in Georgia and Los Angeles. SEER is a population-based registry that collects basic data on variables such as age, race, stage of disease, common breast cancer biomarkers and treatments. A subset of these women enrolled in the ICanCare study \cite{Katz2017}, in which they were surveyed about the care they received and many other factors. The ICanCare study broadly focused on treatment communication and decision-making in patients with favorable breast cancer. Women were also asked about whether they had a family history of breast cancer, and they provided other information related to their risk of being a carrier of genetic variants associated with breast cancer. The survey was completed by 5080 patients and linked to SEER data. In addition, genetic test results corresponding to pathogenic variants were available for some patients. An external company merged the survey responses and SEER clinical data with genetic testing information obtained from four laboratories that tested patients in the study regions and provided a de-identified dataset. More details regarding the combined datasets are provided elsewhere \cite{Kurian2018}.\\
\indent In this paper, we are interested in using data from the ICanCare study to better understand the prevalence of the pathogenic genetic variants in BRCA1 or BRCA2 among women diagnosed with breast cancer in the USA. Women with breast cancer are increasingly taking genetic tests to find out if they have pathogenic variants in important genes. This information can impact the treatments they receive and is relevant for the care of close relatives. The most well-known breast cancer genes are BRCA1 and BRCA2. The prevalence of pathogenic variants in BRCA1 and BRCA2 in the general population is quite low, estimated to be roughly in the 0.2\% to 0.3\% range \cite{Lippi}. Estimates of prevalence of BRCA1/2 pathogenic variants among breast cancer patients vary from country to country (typically around 2\% to 4\%), but can exceed 20\% among breast cancer patients with a positive familial history of breast cancer \cite{Armstrong2019,HU2021,Dorling2021}. Given the practical importance of these genetic variants to patient prognosis and treatment decision-making, there is a great need to better characterize the prevalence of these pathogenic variants in the population of women newly-diagnosed with breast cancer in the USA. Missing data, however, presents a challenge.\\
\indent Genetic test results (including presence/absence of BRCA1/2 mutation) are not available for some patients in the ICanCare study. Amongst the 5080 women 27.5\% had genetic test results, and amongst those with genetic tests 4.66\% had a pathogenic variant in either BRCA1 or BRCA2.  The current recommendation for genetic testing is based on patient age, personal or family history of cancer, known genetic mutation in the family, and tumor characteristics, although there is substantial variability in how much these recommendations are being followed \cite{Katz2018}. Even if genetic testing is offered, patient interest in undergoing genetic testing is influenced by factors such as age, race, education, and stage of disease \cite{Owens2019}. The sample of women who do have genetic test data results within the ICanCare study are very unlikely to be representative of all the women in the ICanCare study or of the population of all women in these two SEER registries, and we expect the estimated prevalence of mutation among women with observed genetic test results to be an over-estimate. More sophisticated strategies are, therefore, needed to address the missing data.\\
\indent Our strategy for handling missingness in BRCA mutation status is to use other available data to multiply impute BRCA status for women with missing values. Then, we can estimate the prevalence of BRCA mutation in the ICanCare study using the multiply imputed data. For the purposes of this paper we will consider a single variable of whether either BRCA1 or BRCA2 has a pathogenic variant. Some key variables that will help inform our imputation of BRCA mutation status are presence of familial history of BRCA mutation, Jewish ancestry, and familial history of breast cancer. Age, race, presence of ER/PR/HER2 mutations, tumor grade, clinical T-stage, presence of lymph node invasion, and presence of bilateral disease may also be informative. Most of these variables had low missingness rates (0\% - 5\%), but HER2 status, family history of cancer and known familial BRCA mutation had higher missingness rates (10\%-21\%). Summary statistics for these variables along with their missingness rates are given in  \textbf{Table C.1} in the \textbf{Supplemental Material}. \\
\indent Standard multiple imputation methods require us to assume that missingness in BRCA mutation status is independent of unobserved information given the observed data. However, this may not be the case. In particular, we may believe that presence of familial history of mutation, familial history of breast cancer, and other variables may strongly impact choices for whether or not a woman undergoes genetic testing. Since these variables also are observed with missingness, the MAR assumption may likely be violated. We see evidence of this dependence in the data. Logistic regression modeling of whether a woman had an available BRCA1/2 test result using data for the 2863 patients with complete covariate information showed an association between missingness and age, race, familial history of either breast, ovarian cancer or sarcoma, familial history of BRCA1/2 mutations, Jewish ancestry, HER2 status, and geographic location. The odds ratios for this logistic regression model are presented in \textbf{Table C.2} in the \textbf{Supplemental Material}. Since these variables are related to missingness in BRCA mutation status and are also occasionally or even often missing themselves, missingness in BRCA status may likely be MNAR.  \\
\indent This MNAR mechanism has a potential to induce bias in resulting estimates of BRCA mutation rates, since these variables are also related to whether or not the BRCA mutation was present. In particular, we ran a logistic regression model on the 874 patients that received a BRCA1/2 test and had complete information for the clinical and demographic factors listed above. In this logistic regression we used the Firth correction to avoid quasi-separation due to the rare outcome. The following variables were clearly associated with having a BRCA1/2 pathogenic variant: age, relatives with history of either breast, ovarian cancer or sarcoma, relatives with known BRCA1/2 mutations and Jewish ancestry(borderline). The odds ratios for this logistic regression model are presented in the first column of  \textbf{Table C.3} in the \textbf{Supplemental Material}. Many of the same variables that are associated with receiving a test are also associated with the positivity rate of the test. Since these variables also have missingness themselves, there is a need to carefully guard against bias due to the MNAR missingness in the multiple imputation process.

We performed sequential regression multiple imputation of the missing data using the {\it mice} program in R using several of the methods explored in this paper. There were four variables with missingness exceeding 10\%: BRCA1/2 test results, family history, known pathogenic variant and HER2.  For these four variables we created response indicators $R_j$  and  offset variables $Z_j$ from \ref{offsetbinary2}, $j=1,..4$.
Multiple imputations were generated using the following three methods: (1) standard SRMI, (2) SRMI-MI and (3) SRMI-Offset.
When using the SRMI-MI method we imputed each variable $j$ in the dataset conditional on all other variables, and all of the above $R_{(-j)}$. 
When using the SRMI-Offset method we imputed BRCA1/2 test, family history, known pathogenic variant, and HER2 status conditional on all other variables and all of the above $Z_{(-j)}$. The rest of the variables were imputed conditional on all other variables and $R_j$.

 Logistic regression models were used for imputing binary variables, and multinomial logistic regression was used for imputing variables with more than 2 categories. We treated clinical stage and tumor grade as categorical. For binary variables with low prevalence, imputation using the `logreg' option in \textit{mice} is unstable and can produce bias in downstream prevalence estimates as shown by simulation in \textbf{Supplemental Material}. To address this issue, we impute BRCA1/2 status using the proposed stratified bootstrap strategy to draw parameters within the imputation algorithm. For each imputation method, we obtain 10 multiple imputations based on sequential regression algorithms that were run for 50 iterations. The marginal prevalence of BRCA1/2 mutation was then estimated, along with corresponding standard errors. \\ 
\indent \textbf{Table \ref{tab:prevalences}} shows the estimated prevalence of BRCA1/2 pathogenic variants from the complete cases and from the different multiple imputation methods. As expected, the multiple imputation methods give lower estimates than the complete case analysis. The extensions of the SRMI that make use of the missing data indicators give slightly lower estimates than obtained from SRMI. Since we believe missingness is MNAR, we would trust results from the SRMI-MI and SRMI-Offset methods over the estimates from SRMI. In \textbf{Table C.3} in the  \textbf{Supplemental Material}, we also present the estimated associations between having a BRCA1/2 pathogenic variant and the various risk factors for each of the imputation strategies. The results compared to the complete case analysis are broadly similar, but there are some differences. Notably, the associations for tumor grade and clinical T-stage are larger than in complete case analysis, and the associations for Jewish ancestry are smaller. As expected, the width of the 95\% confidence intervals for the odds ratios from the multiply imputed datasets tend to be smaller than seen in complete case analysis.

\begin{table}[ht]
\small
\centering
\caption{\label{tab:prevalences}Estimated prevalence of BRCA1/2 pathogenic variants}
\begin{tabular}{lll}
\hline
 & Estimate ($\times$ 100) & Standard Error ($\times$ 100) \\
\hline
Complete cases &  4.66	 &	0.56 \\
SRMI &  2.82	 &	0.47 \\
SRMI-MI &  2.77	 &	0.39 \\
SRMI-Offset &  2.65	 &	0.34 \\

\hline

\end{tabular}
\end{table}

\section{Discussion} \label{discussion}

Standard software for implementing sequential multiple regression imputation (SRMI) assumes that missingness does not depend on unobserved information, called missing at random (MAR). Several researchers have proposed adaptations of existing sequential multiple imputation procedures in settings where missingness is not at random (MNAR) \cite{Tompsett2018, Jolani2012}. For example, Tompsett et al. (2018) proposes handling MNAR missingness by including missing data indicators as predictors in the sequential imputation models. In terms of rigorous statistical justification, however, little work has been done to provide guidance for handling of MNAR missingness within chained equations imputation algorithms in general. \\
\indent In this paper, we provide statistical justification for the missing data indicators method of Tompsett et al. (2018) and propose several extensions that can result in improved performance in terms of bias in the final data analysis. We approach this problem by first deriving the ideal imputation distribution as a function of observed data and assumed models for data missingness, viewing SRMI as an approximation to Bayesian MCMC estimation. Using Taylor series approximations and other methods, we obtain regression model approximations to the ideal imputation distribution to use in practice. We focus our attention on a particular MNAR setting, where missingness for a given variable may depend on \textit{other} variables with missingness. Handling of MNAR missingness in a given variable based on its \textit{own} missing values is a more challenging problem, and we refer the reader to Beesley and Taylor (2021) for recent work in this area \cite{Beesley2021}.\\
\indent Through simulation, we found that inclusion of missingness indicators within sequential imputation algorithms (here, called SRMI-MI) can result in reduced bias in estimating outcome models parameters when missingness is MNAR. The degree of bias reduction will likely depend on the strength of the MNAR missingness and the structure of the missingness model. Although not explored here, inclusion of extra parameters in the imputation models could increase the risk of overfitting and may require larger datasets in order to see good bias reduction properties. In our simulations (datasets of size n=2000), we 
did not see increase in bias or substantial increases in variance when SRMI-MI was applied instead of SRMI when missingness was truly MAR.  \\
\indent In some settings, SRMI-MI produced substantial residual bias. We proposed a variety of extensions to the SRMI-MI approach, including use of spline functions of model predictors, inclusion of interactions, and use of fixed offsets calculated as a function of estimated missingness model parameters. In general, approaches including additional interaction terms tended to result in increased standard errors with some benefit in terms of bias reduction. Of all the regression model approximations, the approach using missingness model-based offsets had the best properties on average across the many simulation settings considered. This may be because this approach is making use of more information from the data, since it involves assuming (and fitting) a model for the probability of missingness for each variable. Since we assume missingness in a given variable is independent of its own missing values, parameters in this missingness model may be identified using the observed data. However, this approach may be more sensitive to misspecification of the missingness model.\\
\indent For comparison, we evaluate the performance of the various SRMI adaptations to imputation using the ``exact" imputation model in \ref{condEqR}. This distribution may only be known up to proportionality, and imputation using this distribution may be complicated in general. In our simulations, this approach (SRMI-Exact) resulted in little or no bias in estimating outcome model parameters.\\
\indent With the exception of the SRMI-Exact method, we tried to restrict our focus to methods that are easily implemented within established sequential imputation software. The methods using the offset do require some non-trivial adaptations of the standard SRMI routine (including fitting of models for covariate missingness within the iterative imputation algorithm), and we provide example code guiding implementation with package \textit{mice} in R.\\
\indent The methods were applied to address potential MNAR missingness in data from the ICanCare study, which consists of a probability-sampled cohort of breast cancer patients identified from two SEER registries \cite{Katz2017}. Weighted analyses using data from the ICanCare study and provided sampling weights can be generalized to the entire SEER registry. The estimated weighted prevalence of a pathogenic variant of BRCA1/2 obtained after SRMI-MI imputation was 2.55\%. The corresponding unweighted estimate was 2.65\%. When handling missing data for survey data, the typical practice is to perform multiple imputation ignoring the weights and then incorporate the weights in the final data analysis. An alternative approach would be to apply a finite Bayesian bootstrap to generate synthetic populations for which imputations can be imputed. Inference is then obtained by extensions of multiple imputation combining rules \cite{Zhou2016}. Future efforts will investigate use of this approach or other alternatives for incorporating weights into the SRMI-MI imputation procedure and extensions developed in this paper.

\section*{Acknowledgments}
Lauren Beesley and Irina Bondarenko are co-first authors of this paper. This research was partially supported by National Institutes of Health grants CA225697 and CA129102.

\bibliography{references}
\bibliographystyle{SageV}

\end{document}